\documentclass[prc,twocolumn,showpacs,showkeys,superscriptaddress,nofootinbib]{revtex4}
\usepackage{graphicx,amssymb,amsmath,multirow}
\def\be{\begin{eqnarray}}
\def\ee{\end{eqnarray}}
\def\bc{\begin{center}}
\def\ec{\end{center}}

\def\rmd{{\rm d}}

\newcommand{\lsim}{\stackrel{\scriptstyle <}{\phantom{}_{\sim}}}

\def\half{{\textstyle \frac12}}
\begin{document}
\title{Strangeness Balance in HADES Experiments and the $\Xi^-$ Enhancement}
\author{E.E.~Kolomeitsev}\email{E.Kolomeitsev@gsi.de}
\affiliation{Department of Physics and Centre of Science and Research, Univerzita Mateja Bela, SK-97401 Bansk\'a Bystrica, Slovakia}
\author{B.~Tom\'a\v{s}ik}
\affiliation{Department of Physics and Centre of Science and Research, Univerzita Mateja Bela, SK-97401 Bansk\'a Bystrica, Slovakia}
\affiliation{Czech Technical University in Prague, FNSPE, CZ-11519 Prague 1, Czech Republic}
\author{D.N.~Voskresensky}
\affiliation{National Research Nuclear University "MEPhI", Kashirskoe Avenue 31, RU-11549, Moscow, Russia}

\begin{abstract}
HADES data on a strangeness production in Ar+KCl collisions at 1.76$A$~GeV are analyzed within a minimal
statistical model. The total negative strangeness content is fixed by the observed $K^+$ multiplicities on
event-by-event basis. Particles with negative strangeness are assumed to remain in chemical equilibrium with
themselves and in thermal equilibrium with the environment until a common freeze-out. Exact strangeness
conservation in each collision event is explicitly preserved. This implies that $\Xi$ baryons can be released
only in events where two or more kaons are produced. An increase of the fireball volume due to application of
a centrality trigger in HADES experiments is taken into account. We find that experimental ratios of
$K^-/K^+$, $\Lambda/K^+$ and $\Sigma/K^+$ can be satisfactorily described provided in-medium potentials are
taken into account. However, the calculated $\Xi^-/\Lambda/K^+$ ratio proves to be significantly smaller
compared to the measured value (8 times lower than the experimental median value and 3 times lower than the
lower error bar). Various scenarios to explain observed  $\Xi$ enhancement are discussed. Arguments are given
in favor of the $\Xi$ production in direct reactions. The rates of the possible production processes are
estimated and compared.
\end{abstract}
\date{\today}
\maketitle

\section{Introduction}

A new stage in the study of a strangeness production in heavy-ion collisions (HICs) has been achieved with the
launch of HADES experiment at Schwerionen Synchrotron (SIS) at GSI in Darmstadt. The HADES collaboration has
undertaken a complete measurement of the particles containing strange quarks in the system Ar+KCl at the
bombarding energy of 1.76$A$~GeV. Production of both open and hidden strangeness was investigated. Results on
kaon and $\phi$ meson productions are reported in~\cite{HADES-KKphi,HADES-Ko}, on hyperons
in~\cite{HADES-Hyperons}. The first observation of the doubly strange hyperon $\Xi^-$ at such a low collision
energy is described in~\cite{HADES-Xi}. These new results are complementary to previous analyses by FOPI and
KaoS collaborations~\cite{Revs}, providing new insights upon strangeness production mechanisms and dynamics in
HICs. Two puzzling observations have been reported so far. The first puzzle is the strong enhancement of the
$\phi$ meson yield~\cite{HADES-KKphi}, description of which, possibly, requires the inclusion of a new type of
catalytic reactions~\cite{KT09}, see also the thorough analysis in~\cite{Kampfer,SchadePhD10}. The second
puzzle is the anomalously large $\Xi/\Lambda$ ratio~\cite{HADES-Xi} exceeding the predictions of the
statistical model~\cite{ABR06} and of the transport code~\cite{Chen04}. Such an excess over predictions of the
statistical model is worrisome since it may signal about a new production mechanism not yet manifested in
other observed particle yields.

In this work we formulate the minimal
statistical model with explicit strangeness conservation on event-by-event basis
and analyze the HADES results on strangeness production within this model.
The model is especially
suitable for analysis of the $\Xi$ baryon yield. Application of a centrality trigger in HADES experiments and
in-medium potentials acting on all particles are incorporated. Various scenarios to explain observed $\Xi$
enhancement are discussed. Arguments are given in favor of the $\Xi$ production in direct reactions. The rates
of the possible production processes are estimated and compared.

The paper has the following structure. In Section~\ref{sec:HADES} we discuss the HADES data and define the
strange particle ratios. Our minimal statistical model for the strangeness production is formulated in detail
in Section~\ref{sec:model}. The in-medium potentials acting on nucleons and strange hadrons are introduced in
Section~\ref{sec:potentials}. In Section~\ref{sec:trigger} we discuss effects of centrality bias induced by
the LVL1 trigger. The final results of our calculations with and without the trigger effect are collected in
Table~\ref{tab:results} and compared with the experimental data. In Section~\ref{sec:discuss} we discuss
possible  mechanisms of the $\Xi$ enhancement. Conclusions are drawn in Section~\ref{sec:conclus}.

\section{HADES data}\label{sec:HADES}

The singly strange particle multiplicities measured by HADES~\cite{HADES-Hyperons} are:
\begin{subequations}
\be
\mathcal{M}_{K^+} &=& (2.8\pm 0.4)\times 10^{-2}\,,
\label{hades-data-1-Kp}
\\
\mathcal{M}_{K^-} &=& (7.1\pm 1.9)  \times 10^{-4}\,,
\label{hades-data-1-Km}
\\
\mathcal{M}_{K^0_S} &=& (1.15\pm 0.14)\times 10^{-2}\,,
\label{hades-data-1-Kos}
\ee\be
\mathcal{M}_{\Lambda+\Sigma^0} &=& (4.09\pm 0.59) \times 10^{-2}.
\label{hades-data-1-L}
\ee
\label{hades-data-1}
\end{subequations}
The doubly strange hyperons, $\Xi^-$, were detected in $\Xi^-\to\Lambda\pi^-\to p \pi^-\pi^-$ channel and the
$\Xi^-$ to $\Lambda+\Sigma^0$ ratio was reported in~\cite{HADES-Xi}
\be
R_{\Xi/\Lambda}={\mathcal{M}_{\Xi^-}}/{\mathcal{M}_{\Lambda+\Sigma^0}}=(5.6\pm 3)\times 10^{-3}.
\label{hades-data}
\ee
Using the measured multiplicities~(\ref{hades-data-1}), (\ref{hades-data}) and the strangeness conservation
one can write the multiplicity of unobserved $\Sigma^++\Sigma^-$ baryons
\be
\mathcal{M}_{\Sigma^+ + \Sigma^-} &=&
\mathcal{M}_{K^+}+\mathcal{M}_{K^0} - \mathcal{M}_{\Lambda+\Sigma^0}
\nonumber\\
&-& \mathcal{M}_{K^-}-\mathcal{M}_{\overline{K}^0}-2\,\mathcal{M}_{\Xi^-} -2\,\mathcal{M}_{\Xi^0}\,.
\ee
Unfortunately, not all isospin species are measured in the experiment. Relying on the conservation of isospin
on the time scale of a heavy-ion collision, we can assume that ratios of isospin partner multiplicities in the
final state of collision reflect  the isospin asymmetry of colliding nuclei. The isospin asymmetry for ArK and
ArCl collisions is characterized by the coefficient $\eta=(A-Z)/Z\simeq 1.14$. Hence we can write
\be
\mathcal{M}_{K^+}/\mathcal{M}_{K^0} \approx \mathcal{M}_{\bar{K}^0}/\mathcal{M}_{K^-}
\approx \mathcal{M}_{\Xi^0}/\mathcal{M}_{\Xi^-} \approx 1/\eta
\label{eta-ratios}
\ee
and  estimate the $\Sigma$ baryon yield as
\be
\mathcal{M}_{\Sigma^+ + \Sigma^-}^{\rm (iso)} &=& (1+\eta)\,\mathcal{M}_{K^+} -
\mathcal{M}_{\Lambda+\Sigma^0}
\nonumber\\
&-&(1+1/\eta)\, \mathcal{M}_{K^-}-2\,(1+1/\eta)\,\mathcal{M}_{\Xi^-}
\nonumber\\
&=& (1.68\pm 0.87) \times 10^{-2}.
\label{hades-sig-iso}
\ee
In Ref.~\cite{HADES-Hyperons} the multiplicity of $\Sigma^++\Sigma^-$ baryons was calculated differently: the
information about neutral kaon ($K_S^0$) production was used explicitly via the relation $ \mathcal{M}_{K^0}=
2\mathcal{M}_{K^0_S} - \mathcal{M}_{\bar{K}^0}$, and also one assumed that $\mathcal{M}_{\bar{K}^0}\approx
\mathcal{M}_{K^-}$ and $\mathcal{M}_{\Xi^0}\approx \mathcal{M}_{\Xi^-}$ as in an isospin symmetric matter.
Then the following relation follows
\be
\mathcal{M}_{\Sigma^+ + \Sigma^-}^{\rm (Hades)} &=& \mathcal{M}_{K^+}+2\, \mathcal{M}_{K_S^0} -
\mathcal{M}_{\Lambda+\Sigma^0}
\nonumber\\
&-& 3\, \mathcal{M}_{K^-}-4\,\mathcal{M}_{\Xi^-}
\nonumber\\
 &=& (0.71\pm 0.61) \times 10^{-2}.
\label{hades-sig-recons}
\ee
Note that in the corresponding Eq.~(4) in Ref.~\cite{HADES-Hyperons} the $\Xi^-$ multiplicity enters with
factor two rather than four as it should, since two isospin species of $\Xi$'s carry two strange quarks each.
The resulting value calculated following mentioned Eq. (4) in~\cite{HADES-Hyperons} is, thus, slightly higher,
$(0.75\pm 0.65) \times 10^{-2}$\, than that given by our Eq.~(\ref{hades-sig-recons}). We should stress that
the $\Sigma^{\pm}$ multiplicity based on the isospin symmetry (\ref{hades-sig-iso}) is almost factor two
larger than that obtained according to Eq.~(\ref{hades-sig-recons}) and that quoted in
Ref.~\cite{HADES-Hyperons}. The origin of this deviation is that the ratio of observed $K^0$ to $K^+$
estimated as $2\,\mathcal{M}_{K_S^0}/\mathcal{M}_{K^+}\simeq 0.82^{+0.25}_{-0.19}$ is significantly smaller
than that given by the isospin asymmetry coefficient of the colliding nuclei, cf. Eq.~(\ref{eta-ratios}),
$\eta\simeq 1.14$, as one would expect. This could indicate that, {\em perhaps, too few $K_S^0$ are observed
in the experiment} or the isospin symmetry is violated in strangeness production reactions. Since we see no
grounds not to belief in the isospin symmetry, we prefer to use the estimate~(\ref{hades-sig-iso}) rather
than~(\ref{hades-sig-recons}). Nevertheless, since the value~(\ref{hades-sig-recons}) is already in use
(actually the one obtained from Eq.~(4) in Ref.~\cite{HADES-Hyperons}), we will compare the results of
statistical model with both values (\ref{hades-sig-iso}) and (\ref{hades-sig-recons}) to see which one is
better described.

Using the data (\ref{hades-data-1-Kp}), (\ref{hades-data-1-Km}), (\ref{hades-data-1-L}) and
(\ref{hades-sig-recons}) we construct the ratios
\begin{subequations}
\begin{align}
&R_{K^-/K^+}={\mathcal{M}_{K^-}}/{\mathcal{M}_{K^+}}=2.54^{+1.21}_{-0.91}\times
10^{-2}\,,
\label{exp-data-KmKp}\\
&R_{\Lambda/K^+}={\mathcal{M}_{\Lambda+\Sigma^0}}/{\mathcal{M}_{K^+}}=1.46^{+0.49}_{-0.37}\,,
\label{exp-data-LKp}\\
&R_{\Sigma/K^+}^{\rm (Hades)}={\mathcal{M}_{\Sigma^++\Sigma^-}^{\rm (Hades)}}/{2\,\mathcal{M}_{K^+}}
=0.13^{+0.15}_{-0.11}\,,
\label{exp-data-SKp}
\end{align}
\label{exp-data}
\end{subequations}
and with the $\Sigma$ yield (\ref{hades-sig-iso}) obtain the ratio
\be
R_{\Sigma/K^+}^{\rm (iso)}={\mathcal{M}_{\Sigma^++\Sigma^-}^{\rm (iso)}}/{2\,\mathcal{M}_{K^+}}=0.30^{+0.23}_{-0.17}\,.
\label{exp-data-SKp-iso}
\ee
As for $\Xi$ production, in order to reduce the dependence of $R_{\Xi/\Lambda}$ on the total strangeness
content of the fireball we will also use the double ratio
\be
R_{\Xi/\Lambda/K^+}={\mathcal{M}_{\Xi^-}}/({\mathcal{M}_{\Lambda+\Sigma^0}\, \mathcal{M}_{K^+}})=0.20^{+0.16}_{-0.12}\,.
\label{hades-data-new}
\ee

\section{Strangeness production within the minimal statistical  model}\label{sec:model}

\subsection{A brief description of the model}
We assume that for a collision at SIS energy the whole energy contained within a nucleus overlap becomes
thermalized. Then, an initially prepared hot and dense nuclear fireball expands in vacuum. Simplifying, we
assume the fireball to be spatially uniform and characterized by a time-dependent temperature $T(t)$, baryon
density $\rho_B (t)$ and volume $V(t)$. At SIS energies the fireball consists mostly of strongly interacting
nucleons, $\Delta$ isobars and pions, cf.~\cite{Voskre-HIC}. It is assumed that the fireball expansion lasts
till a moment of freeze-out characterized by the values $\rho_{B,\rm fo}$ and $T_{\rm fo}$. Henceforth
in-medium particle thermal momentum distributions become distributions of free-streaming particles.

Strange particles and antiparticles need a special care. In the statistical model for the strangeness
production used before in Refs.~\cite{Ko83,KVK95}, the strange particles are assumed to be most efficiently
produced at the early hot and dense stage of the nuclear fireball.

At SIS energies the fireball is baryon-rich, therefore the created kaons ($K^+$ and $K^0$) have the longest
mean free path compared with strangeness $-1$  hadrons. After being produced in some processes together with
particles carrying $s$ quark, they can easily move off the production point, and either leave the fireball
immediately or, first, thermalize via elastic kaon-nucleon scatterings and then leave it at some intermediate
stage. Since the strangeness production probability is very small, even if the kaon stays in the fireball for
a while, there is little chance that it meets anti-kaon or hyperon and is absorbed by them. Thus, in the
course of collision the amount of
negative strangeness of the fireball grows. The accumulated strangeness is
redistributed among $K^-$, $\bar K^0$, $\Lambda$, $\Sigma$ and $\Xi$ baryons\footnote{Particles with a higher
strangeness, such as $\Omega$, and antiparticles contribute very little and can be ignored.} sustaining in the
thermal equilibrium with pions, nucleons and deltas until a common freeze-out. The $K^+$ yield measured in the
experiment can be used to normalize the abundance of negatively strange particles. This differs our model from
the statistical model~\cite{ABR06}, which assumes that $K^+$ and $K^0$ remain in thermal and chemical
equilibrium with other constituents of the fireball till the fireball freeze-out. The latter assumption may
hold at much higher collision energies than at SIS energies, see~\cite{TK05}. For processes with a small
number of produced strange particles the attention ought to be paid to exact strangeness conservation in each
collision event. This means, e.g., that $\Xi$ baryons can be produced only in events involving two and more
kaons.

Modification of strange hadrons in medium should be incorporated, otherwise it would be impossible to describe
measured $K^{-}$ data satisfactorily. In our minimal statistical model, the medium effects will be described
in the mean-field approximation, see Section~\ref{sec:potentials}. More involved p-wave kaon-baryon
interactions considered in~\cite{KVK,KVK95} will be disregarded. Note that the HADES data acquisition system
includes the first-level trigger (LVL1), which selects the most central collisions, requiring more than 16
charged particles to be produced in the collision. The sensitivity of strange particle ratios to this
centrality bias will be studied in Section~\ref{sec:trigger}.

\subsection{Kaon event classes}

Strangeness production is a rare event. Typically the creation of one $s\bar{s}$ pair, from which the
$\bar{s}$ quark leaves the system as the $K^+$ or $K^0$ meson,  occurs only in 2--3\% of collisions. We will
call such events \emph{the single-kaon events}. In these events $\Xi$ baryon cannot be created, because it
needs at least two $s$ quarks to be produced. The much rarer events, in which two $K^+$, or two $K^0$, or
$K^+$ and $K^0$, come out from the fireball we will call \emph{the double-kaon events}. Depending on the
number of $s$ quarks remaining in the fireball various strange hadrons and their combinations can be observed
in the final state. So the chemical equilibrium conditions are different for the events with different numbers
of produced kaons. To proceed further, we divide the totality of events with strangeness production into
classes of events with one, two, three and so on, $s\bar{s}$ quark pairs created. We will call them as
$n$-kaon events and denote the probability of creation of exactly $n$ strange quark pairs as $P_{s\bar s}^{(n)}$. Now
we will show how this probability can be related to the observed multiplicity of $K^+$ mesons
(\ref{hades-data-1-Kp}).

Let $\mathcal{W} (\rho_B,T)$ be the probability of $s\bar{s}$ pair production per unit volume and per unit
time as a function of the temperature and the baryon density. The integral probability of the pair production
is given by
\be
 W =  \int_0^{t_{\rm fo}}V(t){\mathcal{W}}(\rho_B(t),T(t))\,\rmd t\, .
 \ee
The integral is taken over the fireball evolution time until the freeze-out moment $t_{\rm fo}$. The
probability of creation of exactly $n$ pairs (the $n$-kaon event) is determined by the Poisson distribution
\be
P_{s\bar s}^{(n)} = W^n \,e^{-W} /n! \,.
\label{averMn}
\ee
For scale-less hydrodynamic expansion, we can express the current fireball volume through the freeze-out
volume $V_{\rm fo}$ and some scaling function, $V(t)\,/ V_{\rm fo}$; the fireball expansion time can be
expressed as $t_{\rm fo}=\tau V_{\rm fo}^{1/3}$, see~\cite{Russkikh92}. Hence, we can write
\be\label{lamb}
W=\overline{\mathcal{W}}\,\tau\, V_{\rm fo}^{4/3}\equiv \lambda\, V_{\rm fo}^{4/3}\, ,
\ee
where $\tau$ and the averaged probability $\overline{\mathcal{W}}$ are constants. Since the probability of the
strangeness production is small, $W\ll 1$, we may expand the exponent in (\ref{averMn}). Keeping terms up to
the third order we have
\be
 P_{s\bar s}^{(1)} &=& \lambda V_{\rm fo}^{4/3}
-\lambda^2\, V_{\rm fo}^{8/3} +\half\, \lambda^3\, V_{\rm fo}^{4}+O(\lambda^4) \,,
\nonumber\\
 P_{s\bar s}^{(2)} &=& \half\lambda^2\, V_{\rm fo}^{8/3}
-\half\lambda^3\, V_{\rm fo}^{4}+O(\lambda^4) ,
\nonumber\\
P_{s\bar s}^{(3)} &=& {\textstyle\frac{1}{6}} \, \lambda^3\, V_{\rm fo}^{4}+O(\lambda^4).
\label{averM123} \ee
The value  $\lambda$ is fixed by the total $K^+$ multiplicity observed in an inclusive collision. Each of $n$
anti-strange quarks produced in the $n$-kaon event can leave the fireball not only as $K^+$ but also as $K^0$.
Since the isospin composition of the fireball dictates the ratio of $K^0$ to $K^+$ multiplicities
(\ref{eta-ratios}), we can write
\be M_{K^+}^{(n)}=\frac{n}{1+\eta} P_{s\bar s}^{(n)}.
\label{MKpl-n}
\ee
Here $M_{K^+}^{(n)}$ is the multiplicity of $K^+$ mesons produced in $n$-kaon events, i.e., the number of
$K^+$ mesons produced in all $n$-kaon events divided by the total number of events.

The experimentally measured multiplicity of kaons (\ref{hades-data-1-Kp}) is expressed through (\ref{MKpl-n})
using Eq. (\ref{averMn}) as
\be \mathcal{M}_{K^+}=\sum_{n} \langle M_{K^+}^{(n)} \rangle
 =\frac{\langle W \rangle}{1+\eta}.
\label{Mclasses}
\ee
The brackets indicate  that we are dealing with quantities averaged over the collision impact parameter.
This averaging means
\be \langle \dots\rangle= \frac{\int_0^{b_{\rm max}} \rmd b\, b\, (\dots)}{\int_0^{b_{\rm max}} \rmd b\, b}\,,
\label{b-aver} \ee
where the integration over the impact parameter runs from 0 up to the maximal possible value $b_{\rm max}$.
The Ar+KCl collision studied in~\cite{HADES-Xi} is nearly symmetrical with the number of nucleons in each
colliding nucleus $A=39.5$. Thus, simplifying we may take $b_{\rm max}=2\, r_0\,A^{1/3}$ with $r_0\simeq
1.12$~fm. This accuracy is sufficient to calculate averaged characteristics of the fireball.

Since we neglect the nuclear surface effects, the initial temperature and density of the fireball do not
depend on the impact parameter. Then from Eq.~(\ref{Mclasses}) and (\ref{lamb}) we find
\be \lambda=(1+\eta)\, \mathcal{M}_{K^+}/\langle V_{\rm fo}^{4/3} \rangle.
\label{lambda}
\ee
Now the probabilities (\ref{averM123}) averaged over the impact parameter can be expressed through the
experimental $K^+$ multiplicity (\ref{hades-data-1-Kp}) up to $O\big( M_{K^+}^4\big)$ as follows
\be \langle P_{s\bar s}^{(1)}\rangle\!\!\! &=&\!\!\! (1+\eta)\,\mathcal{M}_{K^+}\Big[1 - (1+\eta)\zeta^{(2)}
\mathcal{M}_{K^+}
\nonumber\\
&+&  \half (1+\eta)^2\zeta^{(3)} \mathcal{M}_{K^+}^2\Big],
\label{NS1}\\
\langle P_{s\bar s}^{(2)}\rangle\!\!\! &=&\!\!\! \half(1+\eta)^2 \mathcal{M}_{K^+}^2 \Big[\zeta^{(2)} -
(1+\eta)\zeta^{(3)} \mathcal{M}_{K^+}\Big],
\label{NS2}\\
\langle P_{s\bar s}^{(3)}\rangle\!\!\! &=&\!\!\! {\textstyle \frac16}(1+\eta)^3 \zeta^{(3)} \mathcal{M}_{K^+}^3\,.
\label{NS3}
\ee
Here we introduced the numerical coefficients
\be
 \zeta^{(n)}=
 \big\langle V_{\rm fo}^{\frac43\, n} \big\rangle/
 \big\langle V_{\rm fo}^{4/3} \big\rangle^n\,.
\label{zetaKn}
\ee
The fireball volume at freeze-out can be expressed as
\be
 V_{\rm fo}(b) = 2\,A\,F(b/b_{\rm max})/\rho_{B,{\rm fo}}\,,
\label{Vfo}
\ee
where the  function $F(x)$ describes the overlap of two colliding nuclei. For the overlap function we take the
parameterization from Appendix of Ref.~\cite{Gosset77}, which for the symmetrical collision is
\be
&& F(x)=(x-1)^2 \,[1+(3/ \sqrt{2}-1) x]\,,
 \label{F}\\
&&\int_0^1\rmd x\, x\, F(x)=(1+\sqrt{2})/20\,.
\nonumber \ee
The inclusive fireball volume equals to $\langle V_{\rm fo}\rangle
\approx A/(2\rho_{B,{\rm fo}})$\,. Using Eq.~(\ref{F}) in
averaging (\ref{b-aver}), we obtain from (\ref{zetaKn})
\be \zeta^{(1)}=1\,,\quad \zeta^{(2)}=2.51\,,\quad \zeta^{(3)}=8.11\,.
 \label{zeta123}
 \ee
We see that the volume dependence of the $s\bar{s}$ production probability $W$ leads to an enhancement,
$\zeta^{(n)}>1$, of the multi-pair production probability $\langle P^{(n>1)}_{s\bar{s}}\rangle$.

After all $K^+$ and $K^0$ mesons have left the fireball in the given $n$-kaon event, the fireball becomes
negatively strange with the total strangeness multiplicity
\be M_S^{(n)}=n\,P_{s\bar s}^{(n)}\,.
\label{MS}
\ee
With the help of the experimental kaon multiplicity
(\ref{hades-data-1-Kp}) we estimate
\be
&\langle M_{S}^{(1)}\rangle=5.2\times 10^{-2}\,,\,
\langle M_{S}^{(2)}\rangle=8.8\times 10^{-3}\,,\, &
\nonumber\\
&\langle M_{S}^{(3)}\rangle=8.7\times 10^{-4}\,. &
\ee
Thus, $\langle M_{S}^{(2)}\rangle/(1+\eta)\mathcal{M}_{K^+}\simeq 15\%$ of kaons are produced pairwise, and
$\langle M_{S}^{(3)}\rangle/(1+\eta)\mathcal{M}_{K^+}\simeq 1\%$ of kaons are produced triplewise.

It is interesting to compare the negative strangeness concentrations in the fireball averaged over the impact
parameter for the different classes of events
\be
\rho_S^{(n)}=\big\langle M_S^{(n)}/V_{\rm fo} \big\rangle \,.
\label{rs2}
 \ee
From Eq.~(\ref{averM123}) we have
\begin{subequations}
\be
\rho_S^{(1)}\!\!\! &=&\!\!\! (1+\eta) \frac{\mathcal{M}_{K^+}}{\langle V_{\rm fo}\rangle} \Big[
\tilde\zeta^{(1)} -(1+\eta) \tilde\zeta^{(2)} \mathcal{M}_{K^+}
\nonumber\\
& +&\!\!\! \half (1+\eta)^2\tilde\zeta^{(3)} \mathcal{M}_{K^+}^2 \Big],
\label{ns-1}\\
\rho_S^{(2)}\!\!\! &=&\!\!\! (1+\eta)^2 \frac{\mathcal{M}_{K^+}^2}{\langle V_{\rm fo}\rangle}
\Big[\tilde\zeta^{(2)}\, -(1+\eta)\tilde\zeta^{(3)} \mathcal{M}_{K^+} \Big],
\label{ns-2}\\
\rho_S^{(3)}\!\!\! &=&\!\!\!  \frac12}(1+\eta)^3 \frac{\mathcal{M}_{K^+}^3}{\langle
V_{\rm fo}\rangle}\,{\textstyle\tilde\zeta^{(3)}\,,
\label{ns-3}
\ee \label{scons}
\end{subequations}
with
\be
\tilde{\zeta}^{(n)}=\big\langle V_{\rm fo}^{\frac43\, n-1} \big\rangle \big\langle V_{\rm fo} \big\rangle/
\big\langle V_{\rm fo}^{4/3} \big\rangle^n\,.
 \label{zetaKnt}
 \ee
The numerical values of the coefficients $\tilde{\zeta}^{(n)}$ are
\be
\tilde\zeta^{(1)}=0.693\,,\quad \tilde\zeta^{(2)}=1.04\,,\quad \tilde\zeta^{(3)}=2.85\,.
\label{zetat123}
\ee
Using these values we estimate
\be
\frac{10^3 \,\rho_{S}^{(1)}}{\rho_{B,{\rm fo}}}\simeq \frac{10^{4}
\,\rho_{S}^{(2)}}{\rho_{B,{\rm fo}}}\simeq 1.9\,,\,\, \frac{10^5 \rho_{S}^{(3)}}{\rho_{B,{\rm fo}}} \simeq
1.6\,.
\ee

\subsection{Strangeness statistical probability}

The statistical probability that strangeness will be released at freeze-out in a hadron of type $a$ with the
mass $m_a$ is given by the standard Gibbs' formula
\be
 &&P_a = z_S^{s_a}\,V_{\rm fo}\, p_a= z_S^{s_a}\,V_{\rm fo}\, \nu_a\,
e^{B_a\frac{\mu_{B,{\rm fo}}}{T_{\rm fo}}} f(m_a,T_{\rm fo}),
 \label{h-density}
\\
&& f(m,T) =\!\! \intop\!\!\frac{\rmd^3 p}{(2\,\pi)^3}e^{-\frac{\sqrt{p^2+m^2}}{T}} = \frac{m^2\, T}{2\pi^2}
K_2\left(\frac{m}{T}\right ), \label{f-fun} \ee
where $B_a$ is the baryon number of the hadron, the degeneracy factor $\nu_a$ is determined by
the hadron's spin $I_a$ and isospin $G_a$ as $\nu_a=(2\, I_a+1)\,(2\, G_a+1)$\,, $K_2$ is the
MacDonald function. The baryon chemical potential at freeze-out is determined by
\be
 \mu_{B,{\rm fo}}\!\simeq\! -T_{\rm fo}
\ln\big(4\big[f(m_N,T_{\rm fo})+4f(m_\Delta,T_{\rm fo})\big]/{\rho_{B,{\rm fo}}}\big),
 \label{muB}
 \ee
where $m_N$ is the nucleon mass. $\Delta$ isobars are also treated as stable particles with
the mass $m_\Delta=1232$~MeV, and small contribution of heavier resonances, hyperons and anti-particles is
neglected.

The quantity $z_S$ in (\ref{h-density}) is a normalization factor. It is related to a probability to find one
$s$-quark in the hadron $a$. We assume that this factor is the same for all types of strange hadrons. This is
equivalent to the assumption that all strange hadrons carrying $s$-quarks are in chemical equilibrium. For
multi-strange hadrons $z_S$ enters Eq.~(\ref{h-density}) as $z_S^{s_a}$, where $s_{a}$ is the number of
strange quarks in the hadron. The factor $z_S^{s_a}$ follows from the requirement that the sum of
probabilities for production of different strange species and their combinations, which are allowed in the
finale state, equals to one. The factor $z_S^{s_a}$ depends on how many strange quarks are produced. Hence, it
is different in single-, double- and triple-kaon events. Therefore, we introduce the notation
\be
 P_a^{(n)} = (z_S^{(n)})^{ s_a}\,V_{\rm fo}\, p_a\,,
 \ee
where the superscript $n$ indicates to which class of events this
probability and  $z_S$ factor belong.

Consider now the ensemble of nucleus-nucleus collisions with the fixed impact parameter. In a single-kaon
event one $s$-quark can be released  as $\bar{K}$, $\Lambda$ or $\Sigma$. Hence, the normalization condition
for the probabilities (\ref{h-density}) reads
\be P^{(1)}_{\bar{K}}+P^{(1)}_{\Lambda}+P^{(1)}_{\Sigma} = z_S^{(1)}\,V_{\rm fo}\,
(p_{\bar{K}}+p_{\Lambda}+p_{\Sigma})=1.
\label{strange-cons-1}
\ee
The multiplicity $M_a^{(1)}$ of strange hadrons of type $a=\{\bar{K}, \Lambda, \Sigma\}$ produced in such
single-kaon events is given then by
\be
 M_{a}^{(1)}&=& g_a \,M_S^{(1)}\,P^{(1)}_a= g_a \, M_S^{(1)}\,z_S^{(1)}\, V_{\rm fo}\, p_a\,,
 \ee
where $M_S^{(1)}$ is the multiplicity of strange quarks in the fireball at freeze-out in a single-kaon event
given by (\ref{MS}) and (\ref{NS1}). The isospin factor $g_a$ takes into account the asymmetry in the yields
of particles with various isospin projections induced by the global isospin asymmetry of the collision,
$\eta\neq 1$. It depends on the baryon number, $B_a$, strangeness $s_a$ of the hadron $a$, and its third
component of isospin, $t_{3a}$,
 \be
g_a=\frac{\eta^{-(t_{3a}+(B_a+s_a)/2)}}{\sum_{t'_{3}}\eta^{-(t'_{3a}+(B_a+s_a)/2)}}\,,
 \ee
the sum here is taken over all possible values of $t_{3a}$\,. The combination $t_{3a}+(B_a+s_a)/2$ is, of
course, nothing else than a charge of the hadron $a$. For $\eta=1$ this factor reduces to the standard one
$1/(2\, G_a+1)$\,.

In  double-kaon events there can be $\Xi$ baryons besides all
possible combinations of kaon and hyperon pairs. For double-kaon
events the normalization condition (\ref{strange-cons-1}) is to be
replaced by the following one
\be
 &&\big(P^{(2)}_{\bar{K}}+P^{(2)}_{\Lambda}+P^{(2)}_{\Sigma}\big)^2+ P^{(2)}_{\Xi}
\label{strange-cons-2} \\
&&=  z_S^{(2)2}\,V^2_{\rm fo}\, (p_{\bar{K}}+p_{\Lambda}+p_{\Sigma})^2 + z_S^{(2)2}\,V_{\rm fo}\, p_{\Xi}=1.
\nonumber \ee
In Eq.~(\ref{strange-cons-2}) we assumed validity of the classical statistics for bosons and fermions, i.e.,
the probability of two hadron event, $P_{2h}$ with  $h=\{\bar{K}, \Lambda, \Sigma\}$, is assumed to be equal
to the probability of single hadron event squared, $P_{h}^2$. Quantum effects for bosons and fermions make
$P_{2\bar{K}}>P_{\bar{K}}^2$ and $P_{2\Lambda(\Sigma)}< P_{\Lambda(\Sigma)}^2$, but the differences are tiny
for the temperatures under consideration. The factors 2, which  appear (\ref{strange-cons-2}) at the cross
terms after opening the brackets, reflect the number of combinations with which two $s$ quarks can be released
as a given combination of hadrons, e.g., $\bar{K}\Lambda$, $\bar{K}\Sigma$ and $\Sigma\Lambda$.

The multiplicity of the particle $a$ with one $s$-quark produced in the double-kaon events is equal to
\be
M_a^{(2)} =g_a\, 2\,P_{s\bar{s}}^{(2)}\,P^{(2)}_{a}\sum_{b} P^{(2)}_{b}
 =g_a\, M_{S}^{(2)}\,P^{(2)}_{a}\sum_{b} P^{(2)}_{b}\,
 \ee
where $a,b=\{\bar{K}, \Lambda, \Sigma\}$, and $M_S^{(2)}$ is given by Eqs.~(\ref{MS}) and (\ref{NS2}). We take
here into account that the hadron $a$ can be produced in pair or in various combinations with other strange
hadrons. In both cases the quark combinatoric factor 2 is  due, as we discussed in Eq.~(\ref{strange-cons-2}).
The multiplicity of produced $\Xi$ baryons is
\be
 M_\Xi^{(2)} = g_\Xi\,P_{s\bar{s}}^{(2)}\,\,P^{(2)}_{\Xi} =
\half\,g_\Xi\,M_{S}^{(2)}\,\,P^{(2)}_{\Xi} \,.
 \label{Mk2}
\ee
Note the absence of factor 2 in the first equality.

For completeness now consider a vary rare event class when three $K^+$ mesons are produced. For triple-kaon
events the normalization condition is
\be
&&\big(P^{(3)}_{\bar{K}}+P^{(3)}_{\Lambda}+P^{(3)}_{\Sigma}\big)^3+
3\,P^{(3)}_{\Xi}\,
\big(P^{(3)}_{\bar{K}}+P^{(3)}_{\Lambda}+P^{(3)}_{\Sigma}\big)
\nonumber\\ &&\quad+P^{(3)}_{\Omega} =1= z_S^{(3)3}\,V^3_{\rm
fo}\,(p_{\bar{K}}+p_{\Lambda}+p_{\Sigma})^3 \nonumber\\ &&\quad+
3\,z_S^{(3)3} V^2_{\rm fo}(p_{\bar{K}}+p_{\Lambda}+p_{\Sigma})
p_\Xi +z_S^{(3)3}V_{\rm fo} p_{\Omega}\,. \label{strange-cons-3}
 \ee
Factors 3 appearing in this relation show in how many different ways three $s$ quarks can be distributed
between three hadrons, e.g., $\bar{K}\Lambda\Lambda$,  $\bar{K}\bar{K}\Lambda$, or two hadrons, e.g.,
$\Xi\bar{K}$, $\Xi\Lambda$. The multiplicity of hadrons with one $s$-quark produced in a triple-kaon event is
\be
M_a^{(3)}=g_a\,M_{S}^{(3)}P^{(3)}_a\,
\big[\big(P^{(3)}_{\bar{K}}+P^{(3)}_{\Lambda}+P^{(3)}_{\Sigma}\big)^2 +  P^{(3)}_\Xi \big].
\ee
For hadrons with two $s$-quarks we find
\be
M_\Xi^{(3)}=g_\Xi\,M_{S}^{(3)}\, P^{(3)}_\Xi\,\big(P^{(3)}_{\bar{K}}+P^{(3)}_{\Lambda}+P^{(3)}_{\Sigma}\big)\,,
\label{Mk3}
\ee
and for the $\Omega$ baryon
\be
M_\Omega^{(3)}={\textstyle \frac13}M_{S}^{(3)}\, P^{(3)}_\Omega={\textstyle
\frac13} M_S^{(3)}\, z_S^{(3)3}\, V_{\rm fo}\, p_\Omega\,,
\ee
with $M_{S}^{(3)}$ given by Eqs.~(\ref{MS}) for $n=3$ and (\ref{NS3}).

We can easily write the solutions of Eqs.~(\ref{strange-cons-1}), (\ref{strange-cons-2}) and
(\ref{strange-cons-3}) with respect to $z_{S}^{(n)}$:
\be z_S^{(1)} &=&1/ V_{\rm fo}(p_{\bar K}+p_\Lambda+p_\Sigma)\,,
\label{sfugas-1}\\
z^{(2)2}_S &=& 1/(V_{\rm fo}^2\, (p_{\bar K}+p_\Lambda+p_\Sigma)^2+V_{\rm fo}\, p_\Xi)
\label{sfugas-2}\\
&\approx & z^{(1)2}_S \left[1-\frac{p_\Xi}{V_{\rm fo}\, (p_{\bar K}+p_\Lambda+p_\Sigma)^2}\right],
\nonumber\\
z^{(3)3}_S &\approx& z^{(1)3}_S \left[1-\frac{3\,p_\Xi}{V_{\rm fo}\, (p_{\bar K}+p_\Lambda+p_\Sigma)^2}\right.
\nonumber\\
&-& \left.\frac{p_\Omega}{V_{\rm fo}^2\, (p_{\bar K}+p_\Lambda+p_\Sigma)^3}\right].
\label{sfugas-3} \ee
We keep here only the terms up to the first order in $p_\Xi$ and
$p_\Omega$, which constitute correction contributions in the
square brackets in (\ref{sfugas-2}) and (\ref{sfugas-3}) of the
order of 4\% or less for temperatures and densities of interest,
see below. The neglected higher order terms are still stronger
suppressed since $p_\Xi, p_\Omega\ll p_{\bar
K}+p_\Lambda+p_\Sigma$ and $\langle V_{\rm fo}\rangle(p_{\bar
K}+p_\Lambda+p_\Sigma)>1$.

\subsection{Observables}

Having the normalization factors and the chemical potential from Eq.~(\ref{muB}) at our disposal, we can
calculate the multiplicity ratios (\ref{exp-data}), (\ref{exp-data-SKp-iso}), (\ref{hades-data-new}) as
functions of the freeze-out density and temperature.

Consider, first, the $K^-/K^+$ ratio. Keeping terms up to order $\mathcal{M}_{K^+}$ we include contributions to
the $K^-$ yield from single-kaon and double-kaon events
\be
&&R_{K^-/K^+}=\eta\frac{\langle M_{\bar K}^{(1)}+ M_{\bar K}^{(2)}\rangle }{(1+\eta)\, \mathcal{M}_{K^+}}
=\eta\frac{\langle M_S^{(1)}\, z_S^{(1)}\, V_{\rm fo}\, p_{\bar{K}}\rangle}{(1+\eta)\, \mathcal{M}_{K^+}}
\nonumber\\
&&\quad+ \eta\frac{\langle  M_S^{(2)}\,z_S^{(2)2}\,V_{\rm fo}^2\,
p_{\bar{K}}\,\big(p_\Lambda+p_\Sigma+p_{\bar{K}}\big)\rangle} {(1+\eta)\,\mathcal{M}_{K^+}} . \ee
Note that we deal here with quantities observable in inclusive experiments, therefore,  averaging over the
collision impact parameter is performed following Eq.~(\ref{b-aver}). Using Eqs.~(\ref{NS1}), (\ref{NS2}),
(\ref{sfugas-1}), and (\ref{sfugas-2}) we write
\be
 R_{K^-/K^+} &=& \frac{\langle M_{S}^{(1)}\rangle + \langle M_{S}^{(2)}\rangle} {(1+\eta)\,
\mathcal{M}_{K^+}}\, \frac{\eta \, p_{\bar K}}{p_{\bar K} +
p_\Lambda + p_\Sigma}\, \nonumber\\ &-& \frac{\langle V_{\rm
fo}^{-1}\, M_{S}^{(2)}\rangle}{(1+\eta)\,\mathcal{M}_{K^+}}
\frac{\eta \, p_{\bar K}\, p_\Xi}{(p_{\bar K} + p_\Lambda +
p_\Sigma)^3}\,.
 \label{RKmKpin}
 \ee
The second term in (\ref{RKmKpin}) can be expressed  through $\rho_S^{(2)}$ using Eq.~(\ref{rs2}). With the
help of Eq.~(\ref{ns-2}) the $K^-/K^+$ ratio can be finally cast in the form
\be
 R_{K^-/K^+}=\frac{\eta\, p_{\bar K}}{p_{\bar K} + p_\Lambda + p_\Sigma}\, Y_1\,.
 \label{RKmKp}
 \ee
Here  the auxiliary function $Y_1$ is given by
\be
 Y_1 &=& 1 -\frac{(1+\eta) \mathcal{M}_{K^+}\, \tilde\zeta^{(2)}\, p_\Xi} {\langle V_{\rm fo} \rangle
(p_{\bar K} + p_\Lambda + p_\Sigma)^2}\,.
 \label{Y1}
\ee
The second term in (\ref{Y1}) proves to be small. For values of freeze-out temperatures and densities, which
we exploit, it is of the order of 2\%.

Similar calculations yield
\be
 R_{\Lambda/K^+} &=& \frac{1}{\mathcal{M}_{K^+}} \Big\langle M_\Lambda^{(1)} +M_\Lambda^{(2)} +
\eta\frac{M_\Sigma^{(1)} +M_\Sigma^{(2)}}{\eta^2+\eta+1}\Big\rangle
\nonumber\\
&=& (1+\eta)\frac{p_{\Lambda} +\frac{\eta\, p_{\Sigma}}{\eta^2+\eta+1}}{p_{\bar K} + p_\Lambda + p_\Sigma}\, Y_1\,,
\label{RLKp}\\
R_{\Sigma/K^+} &=& \frac{\eta^2+1}{2(\eta^2+\eta+1)}
\frac{\langle M_\Sigma^{(1)}+M_\Sigma^{(2)}\rangle }{\mathcal{M}_{K^+}}
\nonumber\\
&=& \frac{(\eta^2+1)(\eta+1)}{2(\eta^2+\eta+1)} \frac{p_{\Sigma}}{p_{\bar K} + p_\Lambda + p_\Sigma} \,Y_1 \,.
\label{RSKp} \ee

Now let us consider the double ratio $R_{\Xi/\Lambda/K^+}$. In order to keep the terms of the order
$\mathcal{M}_{K^+}$, as those included above in the coefficient $Y_1$, we should take into account that the
$\Xi$ yield is determined by the double- and triple-kaon events, while single-kaon and double-kaon events
contribute to the $\Lambda$ yield. Using Eqs.~(\ref{Mk2}) and~(\ref{Mk3}) we have
\begin{align}
&R_{\Xi/\Lambda/K^+} = \frac{\frac{\eta}{1+\eta}\langle \big(M_\Xi^{(2)}+M_\Xi^{(3)}\big)\rangle}
{\langle M_\Lambda^{(1)} +M_\Lambda^{(2)} + \eta \frac{M_\Sigma^{(1)} +M_\Sigma^{(2)}}{\eta^2+\eta+1}\rangle
\mathcal{M}_{K^+}}
\nonumber\\
&=\frac{\big\langle \half M_S^{(2)} z_S^{(2)2}  V_{\rm fo} p_\Xi
+ M_S^{(3)} z_S^{(3)3} p_\Xi \big(p_\Lambda + p_\Sigma + p_{\bar{K}}\big)  V^2_{\rm fo}\big\rangle}
{\big(p_{\Lambda}+\frac{\eta p_{\Sigma}}{\eta^2+\eta+1}\big)(1+1/\eta)\,\mathcal{M}_{K^+}}
\nonumber\\
&\times\Big[ \big\langle  M_S^{(1)}\, z_S^{(1)} V_{\rm fo} + M_S^{(2)}\,z_S^{(2)2}
\big(p_\Lambda + p_\Sigma + p_{\bar{K}}\big) V^2_{\rm fo} \big\rangle\Big]^{-1}.
\nonumber
\end{align}
With the help of Eqs.~(\ref{sfugas-1}), (\ref{sfugas-2}),
(\ref{sfugas-3}) and (\ref{NS1}), (\ref{NS2}), (\ref{NS3}) we
obtain
\be
 R_{\Xi/\Lambda/K^+} &=& \frac{p_{\Xi}/(p_{\bar K} + p_\Lambda + p_\Sigma)} {\big(p_\Lambda+\frac{\eta
p_\Sigma}{\eta^2+\eta+1}\big)}
\nonumber\\
&\times& \frac{\eta}{1+\eta} \frac{\langle(\half M_{S}^{(2)} + M_{S}^{(3)})/ V_{\rm fo}\rangle }{ \langle
M_{S}^{(1)} +  M_{S}^{(2)}\rangle \mathcal{M}_{K^+}} \,,
 \label{RXLKp0}
  \ee
where only the leading term in $p_\Xi$ is kept. Using that $\langle(\half M_{S}^{(2)} + M_{S}^{(3)})/ V_{\rm
fo}\rangle=\half \rho_S^{(2)}+\rho_S^{(3)}$, following (\ref{rs2}), and Eqs.~(\ref{scons}), we present
\be
R_{\Xi/\Lambda/K^+} =\eta \frac{ p_{\Xi}/(p_{\bar K} + p_\Lambda + p_\Sigma)} {\langle V_{\rm fo}\rangle
\big(p_\Lambda+\frac{\eta p_\Sigma}{\eta^2+\eta+1}\big)} \,Y_2\,,
 \label{RXLKp}
 \ee
with an auxiliary function
\be
Y_2  = \half\tilde{\zeta}^{(2)}\,.
\label{Y2}
\ee
Taking numerical values from Eq.~(\ref{zetat123}) we estimate
$Y_2\simeq 0.52$. The correction from the terms of the higher
order in $\mathcal{M}_{K^+}$ is below 1\%.

For completeness let us now consider the $\Omega^-$ baryon production. The $\Omega^-$ baryons can be
identified through their weak decays $\Omega^-\to \Lambda K^-$ and $\Omega^-\to \Xi^- \pi^0$ or $\Omega^-\to
\Xi^0 \pi^-$. The first decay mode is most simple for detection. Depending on the detection channel it is
convenient to define the following ratios of the multiplicity of reconstructed $\Omega^-$ baryons to the
total multiplicities of decay products
\be R_{\Omega/\Lambda /K^{-}/K^+} &=& \frac{\mathcal{M}_\Omega}{\mathcal{M}_{\Lambda}\, \mathcal{M}_{K^-}\,
\mathcal{M}_{K^+}},
\label{def:ROLKK}\\
R_{\Omega/\Xi/K^+} &=& \frac{\mathcal{M}_\Omega}{\mathcal{M}_{\Xi^-}\, \mathcal{M}_{K^+}}. \label{def:ROXK}
\ee
Keeping only the leading terms we can write the following expressions for these ratios
\be R_{\Omega/\Lambda/ K^{-}/K^+} &=& \frac{(1+1/\eta)\langle M_\Omega^{(3)}+ M_\Omega^{(4)} \rangle} {\langle
M_\Lambda^{(1)}+ M_\Lambda^{(2)} \rangle \langle M_{\bar{K}}^{(1)}+ M_{\bar{K}}^{(2)} \rangle
\mathcal{M}_{K^+}}
\nonumber\\
&=& \frac{ (1+\eta)^2\, p_\Omega/2\eta } {p_\Lambda p_{\bar K}\, (p_{\bar K}+p_\Lambda+p_\Sigma)\, \langle
V_{\rm fo}\rangle^2} Y_3,
\label{ROLKKp}\\
R_{\Omega/\Xi/K^+} &=& \frac{(1+1/\eta)\langle M_\Omega^{(3)}+ M_\Omega^{(4)} \rangle } {\langle M_\Xi^{(2)} +
M_\Xi^{(3)} \rangle\, \mathcal{M}_{K^+}}
\nonumber\\
&=& \frac{(1+\eta)^2 p_\Omega/2\, \eta}{p_\Xi (p_{\bar K}+p_\Lambda+p_\Sigma)\langle V_{\rm fo}\rangle} Y'_3,
\label{ROXKp} \ee
where we used that $z^{(4)}\approx z^{(1)}$ and introduced notations
\be
Y_3 &=& \frac{\langle ({\textstyle \frac13} M_S^{(3)}+M_S^{(4)})/V_{\rm fo}^2 \rangle} {\langle
M_{S}^{(1)} + M_S^{(2)}\rangle^2(1+\eta) \mathcal{M}_{K^+}}
\nonumber\\
&\approx& \frac{\langle V_{\rm  fo}^2 \rangle\langle V_{\rm  fo} \rangle^2}{6\,\langle V_{\rm fo}^{4/3}
\rangle^3} \approx 0.19 \,,
\label{Y3} \\
Y'_3 &=& \frac{Y_3}{Y_2}\approx \frac{\langle V_{\rm fo}^2\rangle \langle V_{\rm fo}\rangle}{3\,\langle V_{\rm
fo}^{4/3}\rangle \langle V_{\rm fo}^{5/3}\rangle} \approx 0.36 \,.
\label{Y3p}
 \ee

In derivations of Eqs.~(\ref{RKmKp}), (\ref{RLKp}), (\ref{RSKp}), (\ref{RXLKp}), (\ref{ROLKKp}), and
(\ref{ROXKp}) we exploited exact strangeness conservation in each class of events with the $n$ created
$s\bar{s}$ pairs. This constraint leads at the end to the factors $Y_{1,2,3}$ and $Y'_3$. If we put these
$Y$'s equal unity, we recover the results of the conventional canonical statistical approach, where the
strangeness is conserved only on average. In practice the latter means, e.g., that $\Xi$ baryons could come
out in events when only one $K^+$ or $K^0$ meson is produced. We see that the effect of the forced strangeness
conservation is very important and hinders the production of multi-strange hadrons.

In ratios (\ref{RKmKp}), (\ref{RLKp}), (\ref{RSKp}) and (\ref{RXLKp}) all probability densities, $p_a$, are
calculated at the freeze-out moment $t_{\rm fo}$, i.e. at the freeze-out temperature $T_{\rm fo}$ and density
$\rho_{B,{\rm fo}}$. The ratios (\ref{RKmKp}), (\ref{RLKp}, (\ref{RSKp}) are completely determined by the
probability densities given by Eq.~(\ref{h-density}). The ratio $R_{\Xi/\Lambda/K^+}$ is additionally
dependent on the mean fireball freeze-out volume $\langle V_{\rm fo}\rangle$.

\begin{figure*}
\centerline{
\parbox{17cm}{\includegraphics[width=17cm]{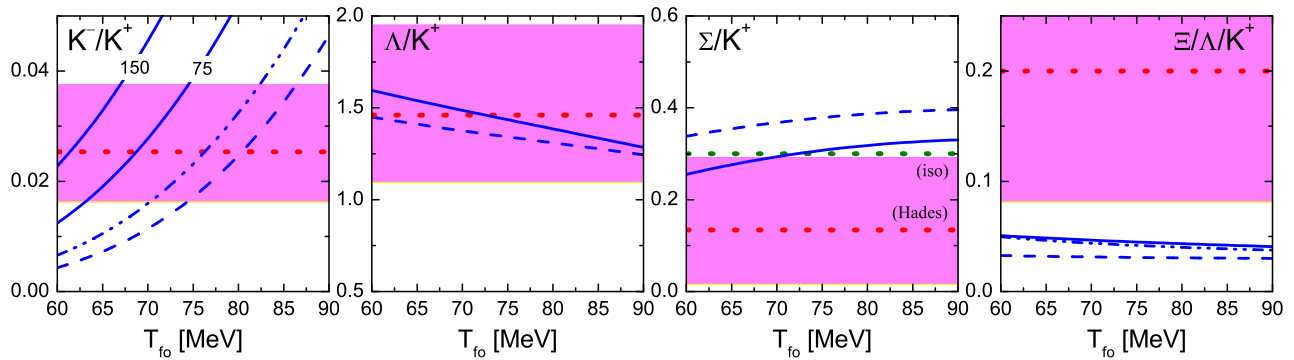}}}
\caption{The ratios (\ref{RKmKp}), (\ref{RLKp}), ( \ref{RSKp}), (\ref{RXLKp}) as functions of the freeze-out
temperature for the freeze-out density $0.6\rho_0$ in comparison with the empirical ratios (\ref{exp-data})
shown by dotted lines. The shaded regions are the experimental error intervals. Dashed curves are calculated
in the absence of in-medium potentials, whereas solid lines, with the scalar and vector potentials. For the
$K^-/K^+$ ratio solid lines labeled as 75 and 150 are computed with $U_{\bar{K}}=-75$ and $-150$~MeV,
respectively. For the ratios involving hyperons solid lines are calculated for $U_{\bar K}=-75$~MeV.
Dash-doubly-dotted lines depict ratios with only nucleon potentials included. On the $\Sigma/K^+$ plot, two
dotted lines show the central values given in (\ref{exp-data-SKp}) and (\ref{exp-data-SKp-iso}) labeled as
``(Hades)'' and ``(iso)'' respectively.}
 \label{fig:statmod}
\end{figure*}

We note that the ratios (\ref{RLKp}), (\ref{RSKp}), (\ref{RXLKp}) depend very weakly on the baryon density.
Indeed, since $p_{\Lambda,\Sigma}\gg p_{\bar K}$, as follows from Eq. (\ref{h-density}), the factor
$e^{\mu_B/T}$ cancels out in Eqs.~(\ref{RLKp}),~(\ref{RSKp}).  The density dependence of the
ratio~(\ref{RXLKp}) is also very weak, since $\langle V_{\rm fo}\rangle \propto 1/\rho_B$, cf.
Eq.~(\ref{Vfo}), and for baryons $p_a \propto e^{\mu_B/T}$, cf. Eq.~(\ref{h-density}), the ratio is
approximately proportional to the combination
\be
\rho_B\, e^{-\mu_B/T}= 4\, f(m_N,T)+16\, f(m_\Delta, T), \ee
which depends only on the temperature and the nucleon and $\Delta$ masses. The density-dependent correction
terms are small as $p_{\bar{K}}/p_{\Lambda,\Sigma}\ll 1$.

In Fig.~\ref{fig:statmod} we depict by dashed lines the strange particle ratios as functions of the freeze-out
temperature. Following Refs.~\cite{KVK95,Voskre-HIC,KV2000} a narrow interval of $\rho_{\rm
fo}=(0.5\mbox{--}0.7)\,\rho_0$, where $\rho_0=0.17$~fm$^{-3}$ is the nuclear saturation density, has been
obtained from the analysis of pion, proton and $K^-$ yields in nucleus-nucleus collisions at SIS and Bevalac
energies. We take here the middle values $\rho_{\rm fo}\simeq 0.6\,\rho_0$. From Fig.~\ref{fig:statmod} we see
that the best description of the $K^-/K^+$ ratio is achieved for $T_{\rm fo}\sim 80$~MeV, whereas for the
$\Lambda/K^+$ we would need a lower temperature $T_{\rm fo}\sim 63$~MeV. The $\Sigma/K^+$ ratio depends weakly
on the temperature and lies 30\% above the upper experimental error-bar of the value (\ref{exp-data-SKp})
obtained using the HADES estimate of the $\Sigma$ multiplicity (\ref{hades-sig-recons}). Compared with the
ratio (\ref{hades-sig-iso}), being reconstructed using only the information on $K^+$ mesons and the isospin
asymmetry of the collision, the calculated $\Sigma/K^+$ ratio lies 30\% above the central value and within the
empirical error bars. The $\Xi/\Lambda/K^+$ ratio is 6.4 times smaller than the central value of the ratio
following from HADES measurements and lies significantly below the lower error bar.

\section{In-medium potentials}\label{sec:potentials}

It is well known that the medium effects are important for description of particle production in HIC at SIS
energies~\cite{Revs,Voskre-HIC}. Particularly, strange particle yields are strongly influenced by
them~\cite{KVK95,LLB97,BratCass,Fuchs06,Aichelin11}. The in-medium modification of the energy spectrum of
particle $a$ is effectively parameterized in terms of scalar $S_a$ and vector $V_a$ potentials
\be
 E_a(p)=\sqrt{m_a^{*2}+p^2}+V_a,\quad m_a^* =m_a+S_a ,
\ee
provided one disregards more involved effects of the $p$-wave interactions. The scalar potential enters the
spectrum through the effective mass $m_a^*$. Description in terms of the $S_a$ and $V_a$ potentials is typical
for relativistic mean-field (RMF) models, cf.~\cite{Fuchs06,KV05}. Inclusion of the potentials $S_a$ and $V_a$
leads to the replacement of the function $f(m_a,T)$ in Eq.~(\ref{h-density}) as
\be
 f(m_a,T)\to f(m_a^*,T)\exp(-V_a/T).
  \ee
For nucleons we use the parameters of the RMF model~\cite{KV05}, $S_N\simeq-190~{\rm MeV}\rho_B/\rho_0$ and
$V_N\simeq + 130~{\rm MeV}\rho_B/\rho_0$, which produce the equation of state  close to the microscopic
Urbana-Argonne equation of state \cite{APR98}. The same potentials are assumed to be valid also for $\Delta$:
$S_\Delta\simeq S_N$, $V_\Delta\simeq V_N$. The vector potentials of hyperons can be related to $V_N$ as
$V_\Lambda=V_\Sigma=2\, V_\Xi=\frac23\, V_N$, according to the number of non-strange quarks in the hyperon.
The scalar potentials follow then as $S_a=[U_a-V_a(\rho_0)]\, \rho_B/\rho_0$, where  the optical potential $U$
acting on a hyperon in an atomic nucleus, $S(\rho_0)+V(\rho_0)=U$, can be constrained  from analysis of
hypernuclei: we take $U_\Lambda=-27$~MeV from~\cite{Hashimoto06}, $U_\Sigma=+24$~MeV from~\cite{Dabrowski99},
$U_\Xi=-14$~MeV from~\cite{Khaustov00}. For $\bar{K}$ mesons we can use the effective scalar potential as in
Ref.~\cite{Kampfer}: $V_{\bar K}=0$, $S_{\bar K}=U_{\bar K}\rho_B/\rho_0$. The $K^-$ optical potential
extracted from kaonic atoms is estimated as $U_{\bar K}=-(70$--150)~MeV. The transport code
calculation~\cite{Kampfer} shows that the $K^-$ production spectra can be described with the potential
$U_{\bar{K}}=-75$~MeV, whereas analysis~\cite{FG07} argues for a stronger attraction, up to  $-150$~MeV.

The ratios~(\ref{RKmKp}), (\ref{RLKp}), (\ref{RSKp}) and (\ref{RXLKp}) calculated with inclusion of the
in-medium potentials are shown in Fig.~\ref{fig:statmod} by solid lines. We use two values of the optical
potential $U_{\bar K}=-75$ and $-150$~MeV. Naturally, the $K^-/K^+$ ratio is the most sensitive to this value.
For hyperons the variation of $U_{\bar K}$ is a small effect and does not change the overall picture.
Therefore, we show only the results for $U_{\bar K}=-75$~MeV advocated in~\cite{Kampfer}. The inclusion of the
attractive potentials leads to an increase of $K^-/K^+$ and $\Lambda/K^+$ ratios, so that one can more easily
find a temperature window when both experimental results are accommodated. The repulsive potential for
$\Sigma$ suppresses the ratio $\Sigma/K^+$ bringing it closer to experimental data. With the inclusion of
in-medium potentials the ratio $\Xi/\Lambda/K^+$ increases. To understand the source of this increase we also
plot the ratio $\Xi/\Lambda/K^+$  with only nucleon potentials included (see the dash-doubly-dotted line).
Comparing solid and dash-doubly-dotted lines we see that increase of the ratio compared to that calculated in
absence of in-medium potentials (dash line) is induced mainly by the nucleon potentials. The latter ones
affect the value of the baryon chemical potential: the factor $\rho_B e^{-\mu_B/T}$ depends now on the baryon
density through the in-medium baryon masses. The $K^-/K^+$ increases also, if nucleon potentials are included
(compare dash-doubly-dotted and dash lines), since the value of $\mu_B$ decreases and the numerator in
Eq.~(\ref{RKmKp}) becomes smaller. However, the main effect is due to the presence of $U_{\bar K}$ (compare
solid and dash-doubly-dotted  lines). The $\Lambda/K^+$ and $\Sigma/K^+$ ratios are insensitive to the nucleon
potentials.

To find the optimal freeze-out temperature we perform a $\chi^2$ fit for ratios $K^-/K^+$, $\Lambda/K^+$ and
$\Sigma/K^+$. Without the in-medium potentials we find $T_{\rm fo}=78.7$~MeV and $\chi^2\simeq 4.6$, if we use
the HADES estimate for $\Sigma^\pm$ yield (\ref{exp-data-SKp}), and we find $T_{\rm fo}=80.0$~MeV and
$\chi^2\simeq 0.35$ using the estimate (\ref{exp-data-SKp-iso}) based on the isospin conservation. With the
potentials (for $U_{\bar K}=-75$~MeV) the fit is improved considerably yielding $\chi^2\simeq 1.7$ and $T_{\rm
fo}=67.1$~MeV for the HADES ratio (\ref{exp-data-SKp}) and $\chi^2\simeq 1.3\times 10^{-2}$ and $T_{\rm
fo}=68.8$~MeV for the isospin ratio (\ref{exp-data-SKp-iso}). The resulting ratios for the last best fit are
collected in Table~\ref{tab:results} in the column labeled ``inclusive''. They should be compared with the
experimental values presented in the column labeled "exp.\ values". We see that the inclusion of in-medium
potentials allows us to reach a reasonable overall agreement with the experiment for the ratios of
singly-strange particles. However, despite the inclusion of the in-medium potentials increases the
$\Xi/\Lambda/K^+$ ratio, this increase is not sufficiently strong to reach even the lower border of the error
bars and the ratio is by the factor 4.2 smaller then the experimental median. As we will see below this
discrepancy increases further by a centrality bias in the HADES experiment, which ought to be taken into
account.

\begin{table}
\renewcommand{\arraystretch}{1.5}
\begin{tabular}{lcll}
\hline\hline ratio             & exp. values & inclusive &
triggered
\\ \hline\hline $(K^-/K^+)\times 10^2$ &  $2.54^{+1.21}_{-0.91}$ &
$2.55$  &  2.55  \\
$\Lambda/K^+$          &  $1.46^{+0.49}_{-0.37}$ & 1.50    &  1.50  \\
$\Sigma/K^+$(Hades)    &  $0.13^{+0.16}_{-0.12}$ & \multirow{2}{*}{0.290}   &  \multirow{2}{*}{0.290}   \\
$\Sigma/K^+$(iso)      &  $0.30^{+0.23}_{-0.17}$ &        &       \\
$\Xi/\Lambda/K^+$      &  $0.20^{+0.16}_{-0.11}$ & 0.047  & 0.026     \\
$(\Omega/\Lambda /K^-/K^+)\times 10^2$ &    ---            & 0.85
&  0.26   \\
$(\Omega/\Xi/K^+)\times 10^2$         &    ---            & 0.42   &  0.23     \\
\hline
\hline
\end{tabular}
\caption{The strange particle ratios (\ref{RKmKp}), (\ref{RLKp}), (\ref{RSKp}), (\ref{RXLKp}), (\ref{ROLKKp}),
and (\ref{ROXKp}) calculated with the inclusion of in-medium potentials (for $U_{\bar K}=-75$~MeV), at the
freeze-out baryon density $\rho_{B, \rm fo}=0.6\,\rho_0$ and  freeze-out temperature $T_{\rm fo}=68.8$~MeV in
comparison with the available experimental values. The columns ``inclusive/triggered'' are results of
calculations with/without inclusion of the LVL1 trigger effects discussed in Section~\ref{sec:trigger}.
 }
\label{tab:results}
\end{table}

\section{Trigger effects}\label{sec:trigger}

In the HADES experiments one uses the LVL1 trigger to select more central
collisions~\cite{HADES-KKphi,HADES-Ko,HADES-Hyperons,HADES-Xi}. The trigger affects  averaging over the impact
parameter, therefore the averaged fireball volume in Eq.~(\ref{RXLKp}) and the numerical factors $\zeta^{(n)}$
and $\tilde\zeta^{(n)}$ change. The trigger effect can be incorporated with the help of an additional weight
function $T_{\rm LVL1}(b)$ embedded in any integration over the impact parameter. There are no direct
experimental tools to constrain this weight function, and we have to rely on a modeling with the help of some
transport code. Obtained from the BUU transport code in~\cite{Kampfer,SchadePhD10}, this function for the
Ar+KCl collisions can be parameterized as
\be T_{\rm LVL1}(b)=\left\{
\begin{array}{cc}
b\,, & b< 3.9~{\rm fm}\\
3.6\, e^{-0.27\, (b/1{\rm fm}-3.75)^2}\,, & b\ge 3.9~{\rm fm}
\end{array}
\right.\,, \ee
(see also Fig.~1 in~\cite{HADES-Hyperons}). Applying the trigger function we should replace the impact
parameter averaging in Eq.~(\ref{b-aver}) as $\langle\dots\rangle\to \langle\dots\rangle_{\rm LVL1}$ with
\be
 \langle\dots \rangle_{\rm LVL1}= {\int_0^{b_{\rm max}}\!\!\!\!\! \rmd b\, T_{\rm LVL1}(b)\, (\dots)}\Big/
\!{\int_0^{b_{\rm max}}\!\!\!\!\!\! \rmd b\, T_{\rm LVL1}(b)}.
\label{b-aver-LVL1}
\ee
The most prominent effect of the multiplicity trigger is an increase of the fireball volume,
\be \langle V_{\rm fo}\rangle_{\rm LVL1}\approx 1.77 \langle V_{\rm fo}\rangle\,.
 \label{VLVL1}
  \ee
This results in a strong decrease of the ratios $R_{\Xi/\Lambda/K^+}$ and $R_{\Omega/\Xi/K^+}$, cf.
Eqs.~(\ref{RXLKp}) and~(\ref{ROXKp}); the ratio $R_{\Omega/\Lambda K^-/K^+}$ decreases still stronger. The
numerical values of the coefficients (\ref{zetaKn}) and (\ref{zetaKnt}) also change as
\be &&\zeta^{(2)}_{\rm LVL1}=1.41\,,\,\, \zeta^{(3)}_{\rm LVL1}=2.39\,,
\label{zetaKnLVL1}\\
&&\tilde\zeta^{(1)}_{\rm LVL1}=0.909\,,\,\, \tilde\zeta^{(2)}_{\rm LVL1}=1.02\,,\,\, \tilde\zeta^{(3)}_{\rm
LVL1}=1.52\,.
\label{zetaKntLVL1}
\ee
The coefficients  $Y_{1,2,3}$ in Eqs.~(\ref{Y1}), (\ref{Y2}), (\ref{Y3}) and (\ref{Y3p}) change slightly and
become equal to $Y_{2,{\rm LVL1}}\simeq 0.51$\, $Y_{3,{\rm LVL1}}\simeq 0.18$, and $Y'_{3,{\rm LVL1}}\simeq
0.35$.

The resulting strange particle ratios are collected in Table~\ref{tab:results} in column ``triggered''. We see
that with the inclusion of the LVL1 trigger effect the statistical model predicts the $\Xi/\Lambda/K^+$ ratio
which is by factor 7.7 smaller than the central value of the HADES measurement and by factor 3.5
smaller than the empirical lower error bar. Thus, the trigger effect pushes the solid, dash-doubly-dotted and
dotted lines shown in Fig.~\ref{fig:statmod} for the $\Xi/\Lambda/K^+$ ratio further downwards.

\section{Discussion}\label{sec:discuss}

As we found above the experimental ratio $\Xi/\Lambda$ measured by HADES cannot be explained within the
minimal statistical model, which is based on strangeness conservation and on the assumption that the
negatively strange particles $\bar{K}$, $\Lambda$, $\Sigma$, and $\Xi$ sustain in thermal equilibrium. The
inclusion of in-medium potentials enlarges the $\Xi$ ratio but not enough to accommodate the experimental
data. Let us discuss now other possible sources of the $\Xi$ enhancement.

(i){\it More attractive $\Xi$ in-medium potential}. One could try to explain the $\Xi$ enhancement by
introduction  of a more attractive $\Xi$ in-medium potential than that we used. Within our model we find that
the value of the potential $U_{\Xi}$ at the saturation nuclear density should be $U_{\Xi}\lsim - 120$~MeV to
increase the ratio $\Xi^-/\Lambda/K^+$ up to the lowest end of the empirical error bar. Such a strong
attraction comparable with the nucleon optical potential is unrealistic. It would imply that $\Xi$ baryon is
bound in nucleus stronger than two $\Lambda$'s, since $2\,(m_\Lambda+U_\Lambda)-(m_\Xi+m_N+U_\Xi+U_N)\sim
100$~MeV$>0$. This would influence the description of doubly strange hypernuclei~\cite{Khaustov00}. The
leading-order analysis of the hyperon and nucleon mass shifts in nuclear matter performed using the chiral
perturbation theory~\cite{SW96} shows that the $\Xi$ shift is much smaller than nucleon and $\Lambda$ shifts.
Recent analysis~\cite{Polinder07,Gasparyan11} confirm the relative smallness of $\Xi N$ scattering lengths.
Nevertheless, for completeness, we should note that there exist some potential models, which predict rather
strong $\Xi N$ interaction, see~\cite{Valcarce10,Kryshen11} and critical discussion in~\cite{Gasparyan11}.

(ii){\it Variations of the freeze-out density}. We could take somewhat larger value of the freeze-out baryon
density. For instance, had we taken $\rho_{B,{\rm fo}}= \rho_{0}$, the calculated ratio $R_{\Xi/\Lambda/K^+}$
would increase but only moderately, from $0.026$ to $0.034$. The latter value is by factor of 3-10 smaller
than the experimental values.

(iii){\it Variations of the $K^+$ multiplicity}. If we vary the $K^+$ multiplicity within the experimental
error bars and take the maximal possible value $\mathcal{M}_{K^+}=3.2\times 10^{-2}$, the ratio
$R_{\Xi/\Lambda/K^+}$ becomes equal to $0.175^{+0.16}_{-0.11}$ instead of the value $0.20^{+0.16}_{-0.11}$
presented in Table.~\ref{tab:results}. The calculated value is still significantly below the experimental
range.

Could it be that the number of produced $K^+$ mesons is underestimated in the experiment? There are two
independent measurements of the $K^+$ yield by KaoS collaboration at the beam energy of 1.8A~GeV  for C+C
collisions~\cite{Laue99} and for Ni+Ni collisions~\cite{Barth97}. If we scale these results down by $A^2$
factor with the corresponding value of $A$ (see argumentation in~\cite{Russkikh92,KT09,KVK95}) we find very
similar results: $\sigma_{\rm C+C}/12^2=(2.1\pm 0.2)\times 10^{-2}$~mb and $\sigma_{\rm Ni+Ni}/58^2=(1.7\pm
0.45)\times 10^{-2}$~mb. Taking the median of 0.02~mb we obtain that for the Ar+KCl collision the total
inclusive $K^+$ production cross section would be equal to $\sigma_{K^+}=31$~mb. Dividing the production cross
section by the geometrical cross section $\sigma_{\rm geom}=\pi\, b_{\rm max}^2$, we estimate the $K^+$
multiplicity for the HADES Ar+KCl experiment as $\mathcal{M}_{K^+}=\sigma_{K^+}/(\sigma_{\rm
geom}\varkappa_{\rm LVL1})$. Here the coefficient $\varkappa_{\rm LVL1}$ takes into account a decrease of the
geometrical cross section because of triggering off peripheral collisions
\be
 \varkappa_{\rm LVL1} &=& \int_0^{b_{\rm max}}\!\!\!\!\! \rmd b\, T_{\rm LVL1}(b)\Big/ \!\int_0^{b_{\rm
max}}\!\!\!\!\!\! \rmd b\, b \simeq 0.45\,.
 \label{kappaLVL1}
 \ee
Thus, we find that the kaon multiplicity expected for the HADES experiment would be
$\mathcal{M}_{K^+}=3.8\times 10^{-2}$, which is larger than the actually observed value and lies beyond the
experimental error bar, cf. Eq.~(\ref{hades-data-1-Kp}). The $\Xi/\Lambda/K^+$ ratio recalculated with such a
kaon multiplicity would be
\be
 R_{\Xi/\Lambda/K^+}=0.14^{+0.10}_{-0.09}\,.
\ee
As we see, the discrepancy between the theory and experiment is not overcame. (The lower experimental limit
$R_{\Xi/\Lambda/K^+}^{\rm exp. min} =0.05$ is still larger than the maximum theoretical estimation
$R_{\Xi/\Lambda/K^+}^{\rm max} \sim 0.03$ obtained by us.)

(iv){\it Earlier freeze-out for $\Xi$s}. The main assumptions of our model are that the negatively strange
sub-system is in thermal equilibrium with a non-strange sub-system (pions, nucleons, $\Delta$s, etc.) and that
negatively strange particles are in chemical equilibrium with each other. The former condition for $\Lambda$
and $\Sigma$ hyperons is supported by the presence of efficient reactions $\Lambda N\leftrightarrow \Lambda
N$, $\Sigma N\leftrightarrow \Sigma N$ and $\Lambda N\leftrightarrow \Sigma N$. The cross sections of these
reactions vary between 80 and 25~mb~\cite{Cugnon90} for the relative momenta between $p_T$ and $2 p_T$, where
$p_T\sim 300$~MeV is the thermal momentum of a baryon at a typical temperature $\sim 70$~MeV. The magnitude of
these cross sections is large enough for the rapid equilibration. For the $\bar{K}$, the thermal equilibrium
is maintained by the reactions $K^-N\leftrightarrow \pi \Lambda(\Sigma)$, which are also responsible for the
chemical equilibration among negatively strange particles. Going through intermediate resonance states
$\Sigma(1385)$, $\Lambda(1405)$, $\Lambda(1520)$, these reactions have large cross sections, which can be even
further enhanced in medium~\cite{Lutz04}. Thus, one may hope that the model assumptions hold at least for the
strangeness $S=-1$ particles.

On the other hand, the $\Xi$-nucleon interaction, as we discussed above, is expected to be smaller than the
$\Lambda(\Sigma)$-nucleon interaction. Calculations of~\cite{Polinder07} show that $\sigma(\Xi^-p\to \Xi^-
p)\sim 15$~mb, $\sigma(\Xi^0p\to \Xi^0 p)\lsim 15$~mb, and $\sigma(\Xi^-p\to \Lambda\Lambda)\lsim 10$~mb.
These values are indeed smaller than those for $\Lambda(\Sigma)N$ reactions. Scattering of $\Xi$'s on pions
for nearly isospin symmetrical matter is also considerably weaker than the $\pi N$ scattering. Indeed,
according to the chiral effective field theory~\cite{Mai2009}, the isospin averaged scattering length
$a_{\pi\Xi}^+=(2\, a_{\pi\Xi}^{(3/2)} + a_{\pi\Xi}^{(1/2)})/3$ is of a sub-leading order in the chiral
expansion, $a_{\pi\Xi}^+\sim O(m_\pi^2)$\,, similarly to the small value of $a_{\pi N}^+$ scattering length.
The numerical value of $a_{\pi N}^+$ derived in~\cite{Mai2009} is $a_{\pi\Xi}^+\sim -(0.02\mbox{--}0.04)$~fm.
In the p-wave, where the $\pi N$ scattering is dominated by the broad spin-3/2 $\Delta$ resonance (the width
is 120~MeV), there is only a narrow $\Xi(1538)$ spin-3/2 resonance with the width 9~MeV, which contribution is
small because of the smallness of the coupling constant. Given arguments show that $\Xi$ baryons are
presumably weaklier coupled to the non-strange system than the strangeness $-1$ particles, thus $\Xi$ baryons
are having a longer mean free path. This motivates us to consider a possibility that $\Xi$ baryons leave the
expanding fireball at somewhat earlier stage than $\bar{K}$s, $\Lambda$s, and $\Sigma$s. As before, we
continue to assume that $\Xi$ baryons are in thermal and chemical equilibrium with the system before they
leave it at temperature $T_{\rm fo}^{(\Xi)}$ and density $\rho_{B,{\rm fo}}^{(\Xi)}$, which are higher than
the values $T_{\rm fo}$ and $\rho_{B,{\rm fo}}$, respectively. If so, the $\Xi/\Lambda/K^+$ would be given by
the following expression
\be
 R^{\rm(non-eq.)}_{\Xi/\Lambda/K^+} =
 \frac{\eta \widetilde{p}_{\Xi} (p_{\bar K} + p_\Lambda + p_\Sigma)/\langle \widetilde{V}_{\rm fo}\rangle}
{(\widetilde{p}_{\bar K} + \widetilde{p}_\Lambda + \widetilde{p}_\Sigma)^2
\big(p_\Lambda+\frac{\eta p_\Sigma}{\eta^2+\eta+1}\big)} \,Y_2\,,
\label{RXLKp-noneq}
 \ee
where all quantities with tilde are calculated at the temperature $T_{\rm fo}^{(\Xi)}$ and density
$\rho_{B,{\rm fo}}^{(\Xi)}$ corresponding to the moment of the new $\Xi$ freeze-out. Quantities without tilde
are calculated as before at the $T_{\rm fo}$ and $\rho_{B,{\rm fo}}$\,.
\begin{figure}
\includegraphics[width=6cm]{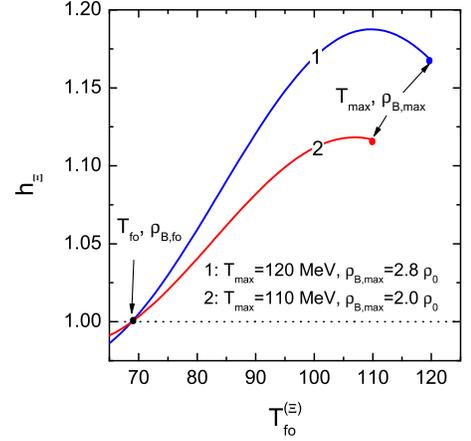}
\caption{
Ratio (\ref{hxi}) plotted as function of the $\Xi$ baryon freeze-out temperature $T_{\rm fo}^{(\Xi)}$. The
corresponding $\Xi$ baryon freeze-out density is calculated according to the polytropic relation
$\rho_{B,{\rm fo}}^{(\Xi)}\propto T_{\rm fo}^{(\Xi)\alpha}$, see text for details. Two curves are drawn for assumed
values of the maximal temperature and density. } \label{fig:hXi-fact}
\end{figure}
Comparing this expression with that in Eq.~(\ref{RXLKp}) we see that the difference is given by the factor
\be
h_\Xi=\frac{R^{\rm(non-eq.)}_{\Xi/\Lambda/K^+}}{R_{\Xi/\Lambda/K^+}}=\frac{\rho_{B,\Xi}}{\rho_{B,{\rm
fo}}}\frac{\tilde{p}_\Xi}{p_\Xi} \frac{(p_{\bar K} + p_\Lambda +
p_\Sigma)^2}{(\widetilde{p}_{\bar K} + \widetilde{p}_\Lambda +
\widetilde{p}_\Sigma)^2} \,.
 \label{hxi}
  \ee
This ratio is plotted in Fig.~\ref{fig:hXi-fact} as a function of the $\Xi$ baryon freeze-out temperature. The
value of the corresponding freeze-out density is evaluated using a polytropic relation between temperature and
density during the fireball expansion $\rho=\rho_{B,{\rm max}}\big(T/T_{\rm max}\big)^\alpha$ with
$\alpha=\log(\rho_{B,{\rm fo}}/\rho_{B,{\rm max}})/\log(T_{\rm fo}/T_{\rm max})$, where $T_{\rm max}$ and
$\rho_{B,{\rm max}}$ are the initial (maximal) temperature and density of the fireball, respectively. For
illustration we use two values for $\alpha$: one, that was advocated by the analysis in Ref.~\cite{Voskre-HIC}
with $T_{\rm max}\simeq 120$~MeV and $\rho_{B{\rm max}}=2.8\,\rho_0$, is shown by line 1, and the other one
with somewhat smaller initial values $T_{\rm max}=110$~MeV and $\rho_{B{\rm max}}=2\,\rho_0$ is shown by line
2.

As we see from Fig.~\ref{fig:hXi-fact}, the assumption of the earlier freeze-out of $\Xi$ baryons may lead
only to a minor enhancement of the $\Xi/\Lambda/K^+$ ratio, $h_\Xi\lsim 1.2$, provided $\Xi$ baryons have
stayed in chemical equilibrium with other strange particles right up to the moment of their freeze-out.

\begin{figure}
\includegraphics[width=5cm]{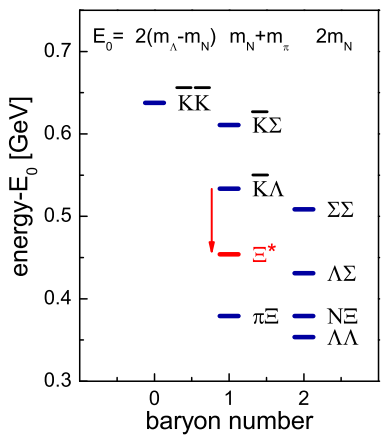}
\caption{Mass spectrum of strangeness --2 states with baryon
numbers 0; 1 and 2 counted from the non-strange ground states with
the energy $E_0$ (shown on top of the figure). }
\label{fig:spectrum}
\end{figure}

(v){\it Direct reactions and sources of $\Xi$ baryons}. From the performed analysis we conclude that to get
any substantial increase in the number of $\Xi$'s we have to assume that these baryons are not absorbed after
being produced and {\em their number is determined by the rate of direct production reactions}, as, e.g., for
di-leptons. This rises up, however, a new question, if there are sufficiently strong sources of $\Xi$ baryons
and enough time to produce the required number of them. There are two types of reactions: {\it strangeness
creation reactions} and {\it strangeness recombination reactions}. The former ones are endothermic and require
a large deposit of kinetic energy from colliding particles
\be \bar{K} N\to K \Xi - { 380~{\rm MeV}}\,,
\nonumber\\
\pi\Sigma \to K \Xi  - { 480~{\rm MeV}}\,,
\nonumber\\
\pi\Lambda \to K \Xi - { 560~{\rm MeV}}\,.
 \label{Screat} \ee
At SIS energies, when the fireball temperature does not exceed $\sim 120$~MeV, these processes have very small
probability. Strangeness recombination reactions are secondary processes involving two strange particles.

In Fig.~\ref{fig:spectrum} we show the mass spectrum of strangeness --2 states in channels with different
baryon numbers $B$. We see that in both single and double baryon channels the states with the $\Xi$ baryon are
on the bottom of the spectrum. Only the double-$\Lambda$ state has a smaller mass. This means that $\Xi$s play
a role of a strangeness --2 reservoir being filled with a decrease of the temperature. The only sink is the
reaction $\Xi N\to \Lambda\Lambda$ having, however, a relatively small cross section~\cite{Polinder07}.

One type of the strangeness recombination reactions is anti-kaon-induced reactions
\be
\overline{K}\Lambda(\Sigma) \to  \Xi\pi + \mathrm{154(232)~{ MeV}}.
\label{Kinduced}
\ee
Their cross sections were calculated in~\cite{LiKo02} and routinely included in transport codes, e.g., see
Ref.~\cite{Chen04}. The cross section has a week dependence on collision energy, of the order of 10~mb. As
suggested in Ref.~\cite{TK11} these reactions can be enhanced in medium because of the attractive potential
acting on $\bar{K}$ mesons. The potential decreases the threshold in the entrance channel of the reaction
(\ref{Kinduced}) so that at some density the threshold drops below the $\Xi^*(1530)$ resonance. The appearance
of the latter narrow resonance above the reaction threshold leads to a strong (but local in energy)
enhancement of the cross section, see Fig.~3 in Ref.~\cite{TK11}. For the choice of the kaon potential
$U_{K^-}(\rho_0)=-75$~MeV~\cite{Kampfer} the drop of the threshold $\bar{K}\Lambda$ for $\rho_B\sim\rho_0$ is
shown in Fig.~\ref{fig:spectrum} by arrow; the $\Xi^*$ resonance is crossed at density $\sim \rho_0$. Thus, at
some stage of the fireball evolution $\Xi$ production in reactions (\ref{Kinduced}) can be stronger than what
was supposed before.

The recombination reactions of the other type proposed in Ref.~\cite{TK11} are the double-hyperon reactions,
\be
\Lambda\, \Lambda \to \Xi N - \mathrm{\phantom{1}26~{ MeV}}\,,
\nonumber\\
\Lambda\, \Sigma \to \Xi N + \mathrm{\phantom{1}52~{ MeV}}\,,
\nonumber\\
\Sigma\, \Sigma \to \Xi N+ \mathrm{130~{ MeV}}\,.
\label{Hypinduced}
\ee
The yields of hyperons are an order of magnitude higher than those of anti-kaons, so we expect a higher
contribution from these processes. The parameterizations of cross sections for some double-hyperon reactions
based on the results of calculations~\cite{Polinder07} can be found in~\cite{TK11}. The
calculations~\cite{Polinder07} include the solution of the coupled-channel Lippmann-Schwinger equation with
the potential constrained by the chiral SU(3) symmetry and parameters fixed by empirical data of
nucleon-nucleon and hyperon-nucleon interactions. Recently the double-hyperon processes have been implemented
in transport code~\cite{Li-Chen-Ko-Lee-12} with the cross sections calculated within the Born approximation,
being factor 5 or more higher than the cross sections presented in Ref.~\cite{Polinder07} depending on a
cutoff parameter employed. With so large cross sections the double-hyperon reactions become the main source of
$\Xi$ baryons and the experimental $\Xi/\Lambda$ ratio could be explained. The large cross section in
Ref.~\cite{Li-Chen-Ko-Lee-12} could be an artifact of the Born approximation and the results of
Ref.~\cite{Polinder07} seem to be more realistic.

Summarizing, it seems to us possible that the enhanced yield of $\Xi^-$ baryons observed by HADES
collaboration can be explained by their production in the direct reactions, provided the in-medium enhancement
of the kaon-induced reactions (\ref{Kinduced}) and new hyperon-induced reactions (\ref{Hypinduced}) are taken
into account.

\section{Conclusions}\label{sec:conclus}

We analyzed the recent HADES data on strangeness production in Ar+KCl collisions at 1.76$A$~GeV in the
framework of the minimal statistical model for strangeness. The latter assumes that the total strange charge
of the fireball created in a certain event is constrained by the number of $K^+$ mesons produced in this even,
and that negatively strange particles remain in thermal and chemical equilibrium  during the fireball
evolution until a common freeze out. Inclusion of realistic in-medium potentials for nucleons and anti-kaons
allows to describe satisfactorily $K^-/K^+$, $\Lambda/K^+$, and $\Sigma/K^+$ ratios. However, the ratio
$\Xi/\Lambda/K^+$ comes out by factor $\sim 3$ smaller than the experimental lower error bar and by factor
$\sim 8$ smaller than the median experimental  value. Two effects contribute to this discrepancy: the
strangeness conservation constraint, which requires that $\Xi$ are created only in the events with two or more
produced $K^+$ or $K^0$ mesons, reduces the $\Xi/\Lambda/K^+$ ratio by factor $\sim 2$ and a centrality bias
due to the LVL1 trigger in the HADES setup leads to a further reduction of the calculated ratio by factor
$\simeq 1.8$, see Eqs.~(\ref{b-aver-LVL1}), (\ref{VLVL1}).

Variation of parameters of the model, such as potentials, freeze-out density and the $K^+$ yield within the
experimental error bars, does not allow to accommodate the data. The assumption, that the created $\Xi$
baryons reach the chemical equilibrium with other strange particles but then leave  the fireball (at their own
freeze-out moment)  earlier than $S=-1$ strange particles, does not allow to produce the sufficient $Xi$
enhancement.

Thus, to overcome the contradiction we suggest that the $\Xi$ baryons do not equilibrate chemically with other
strange particles and the $\Xi$ yield is determined by the direct production reactions. Various $\Xi$
production reactions are discussed.

\begin{acknowledgments}
The study presented here was partially motivated by discussions on workshops organized by the TORIC/TURIC
network. The work was supported by grants VEGA~1/0457/12 and APVV-0050-11 (Slovakia), as well as
MSM~6840770039 (Czech Republic).
\end{acknowledgments}


\end{document}